\documentclass[12pt]{article}

\usepackage{indentfirst}
\usepackage{graphicx, amsmath,amsthm,amsfonts, amssymb}
\usepackage{color, soul, comment}
\usepackage{authblk}
\usepackage{mathtools}

\usepackage{enumerate}

\newtheorem{theorem}{Theorem}

\newtheorem{example}{Example}

\title{Hyperbolic Discounting of the Far-Distant Future}
\author[1]{Nina Anchugina}
\author[2]{Matthew Ryan}
\author[1]{Arkadii Slinko} 
\affil[1]{\small Department of Mathematics, University of Auckland}
\affil[2]{\small School of Economics, Auckland University of Technology}
\affil[ ]{\small \tt n.anchugina@auckland.ac.nz, mryan@aut.ac.nz, a.slinko@auckland.ac.nz}
\date{February 2017}
\begin{document}

\maketitle

\bigskip
\bigskip
\bigskip
\bigskip

\noindent
{\bf Abstract.}  We prove an analogue of Weitzman's \cite{weitzman1998far} famous result that an exponential discounter who is uncertain of the appropriate exponential discount rate should discount the far-distant future using the lowest (i.e., most patient) of the possible discount rates. Our analogous result applies to a hyperbolic discounter who is uncertain about the appropriate hyperbolic discount rate. In this case, the far-distant future should be discounted using the probability-weighted {\em harmonic mean} of the possible hyperbolic discount rates.

\par\bigskip\bigskip
 

\vfill
\noindent

\setcounter{page}{0}
\thispagestyle{empty}
\newpage

\section{Introduction}
Consider an individiual -- or Social Planner -- who ranks streams of outcomes over a continuous and unbounded time horizon $T=[0, \infty)$ using a discounted utility criterion with discount function $D\colon T \to (0, 1]$. We assume throughout that $D$ is differentiable, strictly decreasing and satisfies $D(0)=1$. For example, $D$ might have the \emph{exponential} form \[D(t)=e^{-rt}\] for some constant \emph{discount (or time preference) rate}, $r>0$. Such discounting may be motivated by suitable preference axioms (\cite{Harvey1986}) or as a survival function with constant hazard rate, $r$ (\cite{sozou1998hyperbolic}). For an arbitrary (i.e., not necessarily exponential) discount function, Weitzman (\cite{weitzman1998far}) defines the \emph{local (or instantaneous) discount rate}, $r(t)$, using the relationship:
\begin{equation}
\label{localDR}
D(t)=\exp\left(-\int_{0}^{t}r(\tau)d\tau\right)\ \
\Leftrightarrow\ \ r(t)=-\frac{D'(t)}{D(t)}
\end{equation}
Note that $r(t)$ is constant if (and only if) $D$ has the exponential form. 

Weitzman (\cite{weitzman1998far}) considers a scenario in which the decision-maker is uncertain about the appropriate discount function to use. She may, for example, be uncertain about the true (constant) hazard rate of her survival function, as in  \cite{sozou1998hyperbolic}. The decision-maker entertains $n$ possible scenarios corresponding to $n$ possible discount functions $D_{i}$, $i=1,2,...,n$, with associated local discount rate functions $r_{i}$. Suppose that scenario $i$ has probability
$p_{i}>0$, with $\sum_{i=1}^{n}p_{i}=1$, and that the decision-maker discounts according to the \emph{expected (or certainty equivalent)} discount function
\begin{equation}
\label{mixedDF}
D=\sum_{i=1}^{n}p_{i}D_{i}
\end{equation}
(Such a discount function may also arise if the decision-maker is a utilitarian Social Planner for a population of $n$ heterogeneous individuals, as in  \cite{jackson2015collective}.)

Let $r$ be the local discount rate function associated with certainty equivalent discount function (\ref{mixedDF}).  Weitzman \cite{weitzman1998far} studies the limit behaviour of $r(t)$ as $t\rightarrow\infty$. He proves that if each $r_i(t)$ converges to a limit, then $r(t)$ converges to the lowest of these limits. In other words, if \begin{equation*}
\lim_{t\to \infty} r_i(t)=r_i^*
\end{equation*}
for each $i$, then 
\begin{equation}
\label{limit_rate}
\lim_{t\to \infty} r(t) = \min \{r_1^*, \ldots, r_n^*\}.
\end{equation}

Moreover, if each $r_{i}$ is constant (i.e., each $D_{i}$ is exponential), then $r(t)$ declines \emph{monotonically} to this limit (\cite{weitzman1998far}).\footnote{In fact, this is true more generally -- see \cite[footnote 6]{weitzman1998far}.}

\begin{example}
	\label{eg1}
	Suppose each $D_{i}$ is exponential, so that $r_{i}(t)=r_{i}$ is constant. Then the results in \cite{weitzman1998far} imply that $r(t)$ declines monotonically with limit $\min_i{r_i}$.  
	Figure \ref{graphMixExp2} illustrates for the case $n=3$, $r_1=0.01$, $r_2=0.02$, $r_3=0.03$ and $p_1=p_2=p_3=1/3$.  
	\begin{figure}[h!] 
		\hspace{-1.5cm}
		\includegraphics{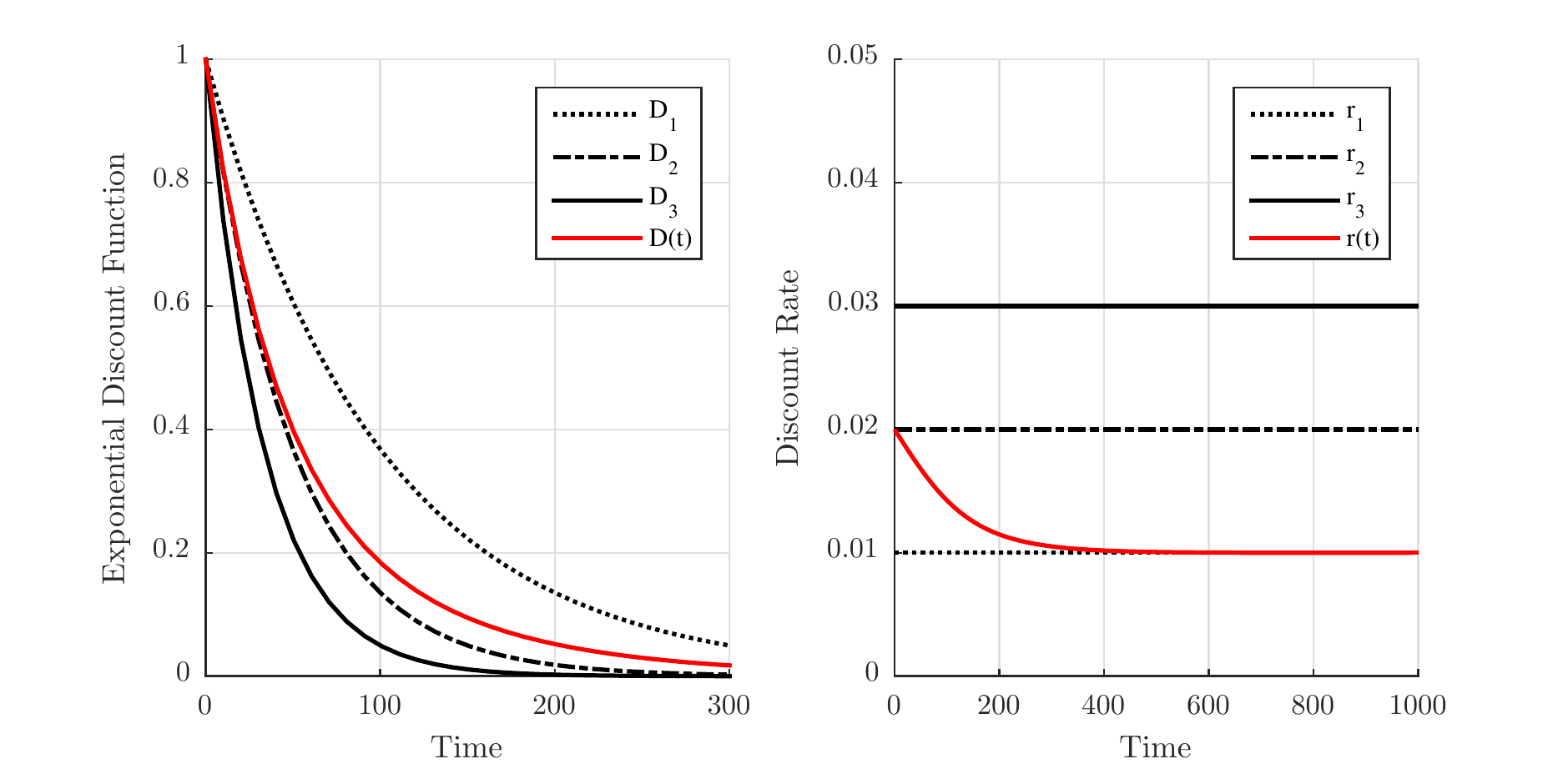}
		\caption{Exponential Discount Functions}
		\label{graphMixExp2}
	\end{figure}
	
\end{example}

Weitzman's result may be interpreted as saying that the mixed discount function (\ref{mixedDF}) behaves locally as an exponential discount function with discount rate (\ref{limit_rate}) when discounting outcomes in the far distant future. This result is most salient if the the individual $D_{i}$ functions are themselves exponential, as in Example \ref{eg1}. However, many individuals do {\em not} discount exponentially (\cite{frederick2002time}). If the $D_{i}$ functions all fall within some non-exponential class, it is natural to characterise the local asymptotic behaviour of (\ref{mixedDF}) using the same class of functions. The next section considers the hyperbolic class.

\section{The case of proportional hyperbolic discounting}

In this section we assume that each $D_{i}$ has the (proportional) hyperbolic form
\[D_i(t)\  \ = \frac{1}{1+h_i t}\]
where $h_i>0$ is the {\em hyperbolic discount rate}. We further assume that $h_1 > h_2 > \ldots > h_n$. In particular, $D_{1}$ exhibits the most ``patience'' and $D_{n}$ the least -- see \cite{ARSaggregation} and Example \ref{eg2}.

Note that 
\[ r_{i}(t)\ =\ -\frac{D_{i}'(t)}{D_{i}(t)}\ =\ \frac{h_{i}}{1+h_{i}t}\]
and hence $r^{*}_{i}=0$ for each $i$. In other words, the limiting local (exponential) discount rate is the same for each discount function, reflecting the fact that hyperbolic functions decline more slowly than exponentials for large $t$. Weitzman's result is not very informative for this scenario.

Instead, we should like to have a local {\em hyperbolic} approximation to the mixed discount function (\ref{mixedDF}), which may not itself have the proportional hyperbolic form. We therefore follow Weitzman's example and define the \emph{local (or instantaneous) hyperbolic discount rate}, $h(t)$, as follows:

\begin{equation}
\label{localHR}
D(t)\ =\ \frac{1}{1+h(t)t}\ \
\Leftrightarrow\ \ h(t)=\left(\frac{1}{D(t)}-1\right)\frac{1}{t}
\end{equation}
Note that $h(t)$ is constant if (and only if) $D$ has the proportional hyperbolic form.
 
{\em How does $h(t)$ behave as $t\to \infty$?} \par\smallskip

The following two results, which are proved in the Appendix, answer this question. In order to state the second result, we remind the reader that the {\em weighted harmonic mean} of non-negative values $x_1, x_2, \ldots, x_n$ with non-negative weights $a_1, a_2, \ldots, a_n$ satisfying $a_1+\ldots+a_n=1$  is
\[
	H(x_1, a_1;\ldots;x_n, a_n)=\left( \sum_{i=1}^n \frac{a_i}{x_i}\right)^{-1}.
\]
It is well-known that the weighted harmonic mean is smaller than the corresponding weighted arithmetic mean (i.e., expected value).

\begin{theorem}
	\label{main1}
	The local hyperbolic discount rate, $h(t)$, is strictly  decreasing.
\end{theorem}

\begin{theorem}
	\label{main2}
	The local hyperbolic discount rate of the certainty equivalent discount function converges to the probability-weighted harmonic mean of the individual hyperbolic discount rates. That is\[h(t)\rightarrow H(h_1, p_1;\ldots;h_n, p_n) \] as $t\rightarrow \infty$.
\end{theorem}

The following example illustrates both results.

\begin{example}
	\label{eg2}
	Suppose $n=3$, $h_1=0.01$, $h_2=0.02$, $h_3=0.03$ and $p_{1}=p_{2}=p_{3}=\frac{1}{3}$. Note that $h_2=0.02$ corresponds to the arithmetic mean of $h_1$, $h_2$ and $h_3$. Figure \ref{gMH} displays the monotonic decline of $h(t)$ towards the weighted harmonic mean $H(h_1, p_1;h_2, p_;h_3, p_3)\approx 0.0164$. 
	
	\begin{figure}[h!] 
		\hspace{-1.5cm}
		\includegraphics{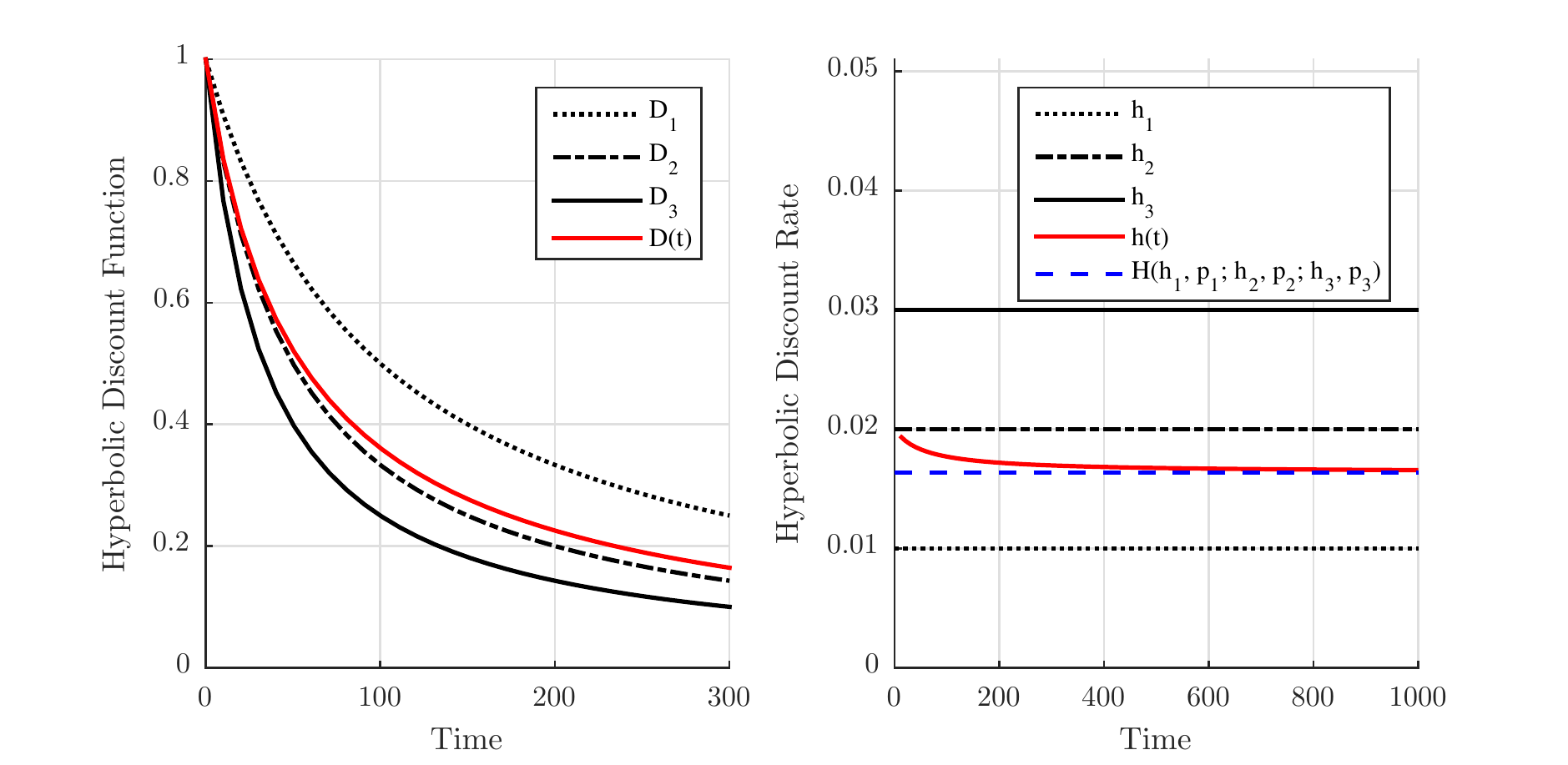}
		\caption{Hyperbolic Discount Functions}
		\label{gMH}
	\end{figure}

\end{example}

\section{Discussion}
With exponential discounting, uncertainty about the (exponential) discount rate implies that the far-distant future is discounted according to the most ``patient'' of the possible discount functions.\footnote{See, in particular, the important reformulation of Weitzman's result by Gollier and Weitzman (\cite{GollierWeitzman2010}), which resolves the so-called ``Weitzman-Gollier puzzle''.} If discounting is hyperbolic, with uncertainty about the (hyperbolic) discount rate, {\em all} possible discount functions matter for the discounting of the far-distant future. The asymptotic local hyperbolic discount rate is, however, below the average (i.e., arithmetic mean) of the possible rates.

\section*{Acknowledgments}
Nina Anchugina gratefully acknowledges financial support from the University of Auckland. Arkadii Slinko was supported by the Royal Society of New Zealand Marsden Fund 3706352.

\appendix
\section{Appendix}
\subsection{Proof of Theorem \ref{main1}}

	We prove this statement by induction on $n$. 
	First we need to prove that the statement holds for $n=2$. In this case:
	\begin{align*}
		h(t) =\left[ \frac{1}{p_1(1+h_1t)^{-1}+p_2(1+h_2t)^{-1}} -1 \right] \frac{1}{t}
	\end{align*}
	for each $t>0$. Rearranging:
	\begin{align*}
		h(t)=\left[ \frac{(1+h_1t)(1+h_2t)}{p_1(1+h_2t)+p_2(1+h_1t)}-1 \right]\frac{1}{t}\ =\ 
		\left[ \frac{1+(h_1+h_2)t+h_1h_2t^2} {p_1 +p_2+(p_1h_2+p_2h_1)t}-1\right]\frac{1}{t}.
	\end{align*}
	Since $p_1+p_2=1$ we obtain:
	\begin{align*}
		h(t)&=\left[ \frac{1+\left(h_1+h_2\right)t+h_1h_2t^2} {1+\left( p_1h_2+p_2h_1\right)t}-1\right]\frac{1}{t}\ =\ 
		\frac{p_1h_1+p_2h_2+h_1h_2t} {1+\left( p_1h_2+p_2h_1\right)t }.
	\end{align*}
	By differentiating $h(t)$:
	\begin{equation} \label{deriv}
		h'(t)=\frac{h_1h_2\left( 1+\left( p_1h_2+p_2h_1 \right)t \right)-\left( p_1h_1+p_2h_2+h_1h_2t\right) \left( p_1h_2+p_2h_1\right)}{ \left[ 1+(p_1h_2+p_2h_1)t \right]^2}
	\end{equation}
	We need to show that $h'(t) < 0$. Since the denominator of \eqref{deriv} is positive, the sign of $h'(t)$ depends on the sign of the numerator. Therefore, we denote the numerator of \eqref{deriv} by $Q$ and analyse it separately:
	\begin{align*}
		Q(t)&\ =\ h_1h_2\left[1+\left(p_1h_2+p_2h_1\right)t\right]-\left(p_1h_1+p_2h_2+h_1h_2t\right)\left(p_1h_2+p_2h_1\right)\\
		&\ =\ h_1h_2-\left(p_1h_1+p_2h_2\right)\left(p_1h_2+p_2h_1\right).
	\end{align*}
	By expanding the brackets and using the fact that $p_1+p_2=1$ implies $1-p_1^2-p_2^2=2p_1p_2$ expression $Q$ can be simplified further:
	\begin{align*}
		Q(t)&\ =\ h_1h_2-p_1^2h_1h_2-p_1p_2h_1^2-p_1p_2h_2^2-p_2^2h_1h_2  \\&\ =\ h_1h_2(1-p_1^2-p_2^2)-p_1p_2(h_1^2+h_2^2)\\&\ =\ -p_1p_2(h_1-h_2)^2.
	\end{align*}
	Therefore, since $h_1\neq h_2$ we have $Q<0$. Hence it follows that $h'(t)<0$ and $h(t)$ is strictly decreasing.
	
	Suppose that the proposition holds for $n=k$. We need to show that it also holds for $n=k+1$.
	When $n=k+1$ we have:
	\[
		D\ =\ \sum_{i=1}^{k+1}p_i D_i\ =\ \left( 1-p_{k+1}\right)\left(\sum_{i=1}^{k}\frac{p_i}{1-p_{k+1}} D_i\right)+p_{k+1}D_{k+1}.
	\]
	Since \[\sum_{i=1}^{k}\frac{p_i}{1-p_{k+1}}=1,\] we may write \[D\ =\  \left( 1-p_{k+1}\right)D^{(k)}+p_{k+1}D_{k+1}.
	\] where \[D^{(k)}\ =\ \sum_{i=1}^{k}\frac{p_i}{1-p_{k+1}} D_i.\] By the induction hypothesis it follows that 
	\[D^{(k)}=\frac{1}{1+h^{(k)}(t)t},\] where $h^{(k)}$ is strictly decreasing.
	Therefore,
	\begin{align*}
		h(t)&=\left[\frac{1}{(1-p_{k+1})D^{(k)}+p_{k+1}D_{k+1}}-1\right]\frac{1}{t}\\
		&=\left[\frac{1}{\left( 1-p_{k+1}\right)\left( 1+h^{(k)}(t)t\right)^{-1}+p_{k+1}\left(1+h_{k+1}t\right)^{-1}}-1\right]\frac{1}{t}.
	\end{align*}
	
	Let $\hat{p_1}=1-p_{k+1}$, $\hat{p_2}=p_{k+1}$, $\hat{h_1}(t)=h^{(k)}(t)$ and $\hat{h_2}=h_{k+1}$.
	Then we have
	\begin{equation*}
		h(t)=\left[ \frac{1}{\hat{p_1}(1+\hat{h_1}(t)t)^{-1}+\hat{p_2}(1+\hat{h_2}t)^{-1}}-1\right]\frac{1}{t}.
	\end{equation*}
	Analogously to the case $n=2$, this expression can be rearranged to give:
	\begin{equation*}
		h(t)=\frac{\hat{p_1}\hat{h_1}(t)+\hat{p_2}\hat{h_2}+\hat{h_1}(t)\hat{h_2}t}{1+\hat{p_1}\hat{h_2}t+\hat{p_2}\hat{h_1}(t)t}.
	\end{equation*}
	However, in contrast to the case $n=2$, $\hat{h_1}$ is now a function of $t$. Thus:
	\begin{equation}
	\label{ht}
		h'(t)\ =\ \frac{N(t)
		}{\left[1+\hat{p_1}\hat{h_2}t+\hat{p_2}\hat{h_1}(t)t\right]^2}.
	\end{equation}
	where
	\begin{align*}
		N(t)=&\left( \hat{p_1}\hat{h}'_1(t)+\hat{h_1}(t)\hat{h_2}+\hat{h}'_1(t)\hat{h_2}t\right) \left( 1+\hat{p_1}\hat{h_2}t+\hat{p_2}\hat{h_1}(t)t\right)\\
		&-\left(\hat{p_1}\hat{h_1}(t)+\hat{p_2}\hat{h_2}+\hat{h_1}(t)\hat{h_2}t\right)\left(\hat{p_1}\hat{h_2}+\hat{p_2}\hat{h_1}(t)+\hat{p_2}\hat{h}'_1(t) t\right).
	\end{align*}
	The denominator of (\ref{ht}) is strictly positive, so the sign of the derivative is the same as that of $N(t)$. Note that
	\[
		N(t)\ =\ \hat{Q}\left(  t\right)  +\hat{h}_{1}^{\prime}(t)\left[  \left(  \hat
		{p}_{1}+\hat{h}_{2}t\right)\left(1+\hat{p}_{1}\hat{h}_{2}t+\hat{p}_{2}\hat{h}_{1}(t)t\right)-\hat{p}_{2}t\left(  \hat{p}_{1}\hat{h}_{1}(t)+\hat{p}_{2}\hat{h}_{2}+\hat{h}_{1}(t)\hat{h}_{2}t\right)  \right]
	\]
	where $\hat{Q}\left(  t\right)  $ is defined as above,
	but with $h_{1}=\hat{h}_{1}\left(  t\right)  $ and $h_{2}=\hat{h}_{2}$. Since
	$\hat{Q}\left(  t\right)  \leq0$ (with equality
	if and only if $\hat{h_1}\left(  t\right)  =h_{2}$) and $\hat{h}_{1}^{\prime}<0$, it
	suffices to show
	\begin{equation}\label{eqn:ineq}
		\left(  \hat{p}_{1}+\hat{h}_{2}t\right)  \left(  1+\hat{p}_{1}\hat{h}%
		_{2}t+\hat{p}_{2}\hat{h}_{1}(t)t\right)  -\hat{p}_{2}t\left(  \hat{p}_{1}\hat
		{h}_{1}(t)+\hat{p}_{2}\hat{h}_{2}+\hat{h}_{1}(t)\hat{h}_{2}t\right)  >0
	\end{equation}
	Cancelling terms on the left-hand side of \eqref{eqn:ineq} leaves us with:
	\[
		\hat{p}_{1}\left(  1+\hat{p}_{1}\hat{h}_{2}t\right)  +\hat{h}_{2}t\left(
		1+\hat{p}_{1}\hat{h}_{2}t\right)  -\left(  \hat{p}_{2}\right)  ^{2}\hat{h}
		_{2}t.
	\]
	We now use the fact that $\left(  \hat{p}_{2}\right)  ^{2}=\left(  1-\hat
	{p}_{1}\right)  ^{2}=1-2\hat{p}_{1}+\left(  \hat{p}_{1}\right)  ^{2}$ to get
	\begin{equation*}
		\hat{p}_{1}\left(  1+\hat{p}_{1}\hat{h}_{2}t\right)  +\hat{h}_{2}t\left(
		1+\hat{p}_{1}\hat{h}_{2}t\right)  -\left[  1-2\hat{p}_{1}+\left(  \hat{p}
		_{1}\right)  ^{2}\right]  \hat{h}_{2}t\
		\ =\ \hat{p}_{1}+\left(  t\hat{h}_{2}\right)  ^{2}\hat{p}_{1}+2\hat{p}_{1}
		\hat{h}_{2}t,
	\end{equation*}
	which is strictly positive. This establishes the required inequality (\ref{eqn:ineq}) and completes the proof.

\subsection{Proof of Theorem \ref{main2}}

	We note that
	\[
	\frac{p_i}{1+h_it} \ =\  \frac{p_i}{h_it}+\epsilon_i(t), 
	\]
	where $\epsilon_i(t)/t^2 \to 0$ when $t\to\infty$. Let $\epsilon(t)=\epsilon_1(t)+\ldots+\epsilon_n(t)$. Hence it follows that:
	\begin{align*}
	\frac{1}{1+h(t)t} \ =\  \sum_{i=1}^n p_iD_i(t) & \ =\  \frac{p_1}{1+h_1 t}+\ldots+\frac{p_n}{1+h_n t}\\ & \ =\   \frac{p_1}{h_1 t}+\ldots+\frac{p_n}{h_n t} +\epsilon(t)
	\\ &\ =\ \biggl( \frac{p_1}{h_1}+\ldots+\frac{p_n}{h_n}\biggr ) \frac{1}{t} +\epsilon(t)\\ &\  =\  \frac{1}{H(h_1, p_1;\ldots;h_n, p_n) t}+\epsilon(t)\\
	&\ =\   \frac{1}{1+H(h_1, p_1;\ldots;h_n, p_n) t}+\hat{\epsilon}(t),
	\end{align*}
	where $\hat{\epsilon}(t)/t^2 \to 0$ as $t\to\infty$. This implies that $h(t) \to H(h_1, p_1;\ldots;h_n, p_n)$ as $t\to\infty$.

\bibliographystyle{plain}
\bibliography{Biblio} 

\end{document}